\documentclass{elsart2}
\usepackage{amssymb}

\begin{document}
\begin{frontmatter}

\title{On elliptic solutions of the quintic
complex one-dimensional Ginzburg--Landau equation}

\author{S.~Yu.~Vernov}

\address{Skobeltsyn Institute of Nuclear Physics, Moscow State
University,\\ Vorob'evy Gory,  Moscow, 119992, Russia}

\ead{svernov@theory.sinp.msu.ru}

\begin{abstract}
The Conte--Musette method has been modified for the search of only elliptic
solutions to systems of differential equations. A key idea of this a priory
restriction is to simplify calculations by means of the use of a few Laurent
series solutions instead of one and the use of the residue theorem. The
application of our approach to the quintic complex one-dimensional
Ginzburg--Landau equation (CGLE5) allows us to find elliptic solutions in the
wave form. We also find restrictions on coefficients, which are necessary
conditions for the existence of elliptic solutions for the CGLE5. Using the
investigation of the CGLE5 as an example, we demonstrate that to find elliptic
solutions the analysis of a system of differential equations is more
preferable than the analysis of the equivalent single differential equation.
\end{abstract}


\end{frontmatter}

\section{Introduction}

At present methods for construction of special solutions of nonintegrable
systems in terms of elementary (more precisely, degenerated elliptic) and
elliptic functions are actively
developed~\cite{Akhmediev,CoMu92,CoMu93,CoMu04,Fan02,Fan03,Pavlov,Kudryashov89,Kudryashov1,Kudryashov2,Kudryashov6,Conte93,CoMu03,Nikitin,santos,Timosh99,VeTish04,VernovJNMP,VladKut}
(see also~\cite{Kudryashov} and references therein). Some of these methods are
intended for the search for elliptic solutions only~\cite{Kudryashov1}, others
allow to find either solutions in terms of elementary functions
only~\cite{CoMu92,CoMu93,Conte93,VladKut} or both types of
solutions~\cite{Akhmediev,CoMu04,Fan02,Fan03,Pavlov,Kudryashov89,Kudryashov2,Kudryashov6,CoMu03,Nikitin,santos,Timosh99,VeTish04,VernovJNMP}.
Note that the methods~\cite{Timosh99,Kudryashov6} allow to find multivalued
solutions as well.

Elliptic and degenerate elliptic functions are single-valued functions,
therefore, such solutions of a nonintegrable system exist only if there exist
the Laurent series solutions of it. Such local solutions can be constructed by
means of the Ablowitz--Ramani--Segur algorithm of the Painlev\'e
test~\cite{ARS} (see also~\cite{HonePainleve,Kudryashov,Tabor}). Moreover for
a wide class of dynamical systems using this method one can find all possible
Laurent series expansions of solutions.  In this way one obtains solutions
only as formal series, that is sufficient, because really only a finite number
of coefficients of these series is used. Examples of construction of such
solutions are given in~\cite{Conte93,VernovTMF}. The Laurent series solutions
give the information about the global behavior of differential systems and
assist to look for both exact solutions~\cite{CoMu03} and  the first
integrals~\cite{Kudryashov04}. The Laurent series solutions can be used to
prove the nonexistence of elliptic solutions~\cite{Hone05,VernovCGLE}  as
well.

In~\cite{CoMu03} R. Conte and M. Musette have proposed a new method for
construction of single-valued special solutions of nonintegrable differential
equations. A key idea of this method is the use of the Laurent series
solutions to transform the initial differential equation into a nonlinear
system of algebraic equations. Using this method one can in principal find all
elliptic and degenerate elliptic solutions. Unfortunately if the initial
differential equation includes the large number of numeric parameters, then it
is difficult to solve the obtained nonlinear system of algebraic equations.

The goal of this paper is to propose a modification of the Conte--Musette
 method, which allows to seek elliptic solutions only. We show that in this case
 it is possible to fix some parameters of
the initial differential system and therefore simplify the resulting system of
algebraic equations. To do this we use the Hone's method, which has been
proposed to prove the non-existence of elliptic solutions~\cite{Hone05}. Note
that using our approach one can find in principal all elliptic solutions.

In~\cite{CoMu03} the authors have transformed the initial system of two
coupled ordinary differential equations into the equivalent single
differential equation and only after this have constructed the Laurent series
solutions. In our paper we demonstrate that the analysis of the system of
differential equations may be more useful than the analysis of the equivalent
differential equation. Moreover in this paper we show that if a system of
differential equations includes a few functions it is possible to find the
analytic form of a function which satisfies this system even without knowledge
of other functions in the analytic form and without elimination of them from
the system.

\section{The Conte--Musette method for the system of differential equations}

In~\cite{CoMu03} R. Conte and M. Musette have proposed a way for searching of
elliptic and degenerate elliptic solutions to a polynomial autonomous
differential equation. In this section we reformulate this method for a system
of such equations:
\begin{equation}
\label{f}
   F_i(\mathbf{\vec{y}}_{;t}^{(n)},\mathbf{\vec{y}}_{;t}^{(n-1)},\dots,
   \mathbf{\vec{y}}_{;t},\mathbf{\vec{y}})=0, \qquad i=1,\dots, N,
\end{equation}
where $\mathbf{\vec{y}}=\{y_1(t),y_2(t),\dots,y_L(t)\}$ and
$y_{j;t}^{(k)}=\frac{\mathrm{d}^ky_j}{\mathrm{d}t^k}$.

It is known that any elliptic function (including degenerate one) is a
solution of some first order polynomial autonomous differential equation. The
classical results of P. Painlev\'e, L. von Fuchs, C.A.A. Briot and J.-C.
Bouquet allow one to construct the suitable form of such an equation, whose
general solution is a meromorphic function with poles of order $p$~(see
details in~\cite{CoMu03})\footnote{The summation in (\ref{subequ}) runs over
nonnegative integer $j$ that are less than or equal to $(p + 1)(m -k)/p$}:
\begin{equation}
\label{subequ}
\sum_{k=0}^{m} \sum_{j=0}^{(p+1)(m-k)/p}h_{j,k} y^j {y_t}^k=0,\qquad
h_{0,m}=1,
\end{equation}
in which $m$ is a positive integer number and $h_{j,k}^{\vphantom{27}}$ are
constants to be determined. The general solution of $(\ref{subequ})$ is either
an elliptic function,  or a rational function of $e^{\gamma x}$, $\gamma$
being some constant, or a rational function of $x$. Note that the third case
is a degeneracy of the second one, which in turn is a degenerate case of the
first one.

The Conte--Musette algorithm is the following~\cite{CoMu03} (see
also~\cite{CoMu04}):

\begin{enumerate}

\item Choose a positive integer number $m$, define the form of
Eq.~(\ref{subequ}) and calculate the number of unknown coefficients $h_{j,k}$.

\item Construct solutions of system (\ref{f}) in the form of the Laurent
series. If such solutions do not exist or they correspond to known exact
solutions, then no unknown single-valued solutions exist. Note that, since
system (\ref{f}) is autonomous, the coefficients of the Laurent series do not
depend on the position of the singular point. They may depend on values of the
numerical parameters of (\ref{f}). In addition, some of these coefficients
(the number of which is less than the order of system (\ref{f})) may take
arbitrary values and have to be considered as new numerical parameters. One
should compute more coefficients of the Laurent series than the number of
numerical parameters in the Laurent series plus the number of $h_{j,k}$.

\item Choose a Laurent series expansion for some function $y_k$ and substitute
the obtained Laurent series coefficients into Eq.~(\ref{subequ}). This
substitution transforms (\ref{subequ}) into a linear and over\-determined
system in $h_{j,k}^{\vphantom{27}}$ with coefficients depending on numerical
parameters.

\item  Eliminate coefficients $h_{j,k}^{\vphantom{27}}$ and get a nonlinear
system in parameters.

\item Solve the obtained nonlinear system.
\end{enumerate}

R.~Conte and M.~Musette note that a computer algebra package is highly
recommended for using of their method~\cite{CoMu04}. Steps of this algorithm
can be implemented in computer algebra systems separately.

For the given system it is easy to calculate the Laurent series solutions to
any accuracy. These computations base on the Painlev\'e test, which has been
implemented in the most popular computer algebra
systems~\cite{Baldwin,Renner,Scheen,XuLi}. Note that when one has computed a
sufficient number of the Laurent series coefficients he can forget about the
system of differential equations and work only with coefficients of the
obtained series. The first package of computer algebra procedures, which
realize the third and the fourth steps of the algorithm, has been written in
AMP~\cite{AMP} by R.~Conte. One can also use our Maple and REDUCE packages of
procedures, which are accessible in Internet~\cite{Vernovsite} and are
described in~\cite{VernovCASC04,VernovCASC05}. So, one passes the first four
steps of algorithm without any difficulties.

At the fifth (last) step one should solve an overdetermined system of
nonlinear algebraic equations. The standard method for solving of such systems
is the construction of a lexicographically ordered Gr\"obner
basis~\cite{Davenport}. The Buchberger
algorithm~\cite{Buchberger,Buchberger2,Davenport} guarantees a construction of
a Gr\"obner basis in a finite number of steps\footnote{An alternative method
for constructing Gr\"obner bases is the involutive
algorithm~\cite{Gerdt1,Gerdt3}.}. However, appropriate estimates of the number
of steps required for constructing of this basis do not exist. The required
computer memory depends, in the general case, exponentially on the number of
the unknowns. Therefore, this number should be made as small as possible.

The purpose of this paper is to show that we can essentially simplify the
algebraic system of equations, which we have to solve on the last step of the
Conte--Musette method, if we search the elliptic solutions only.

In~\cite{CoMu03} R. Conte and M.~Musette have used their method to find wave
solutions of the complex cubic Ginzburg--Landau equation (CGLE3). The
nonexistence of elliptic travelling and standing wave solutions of the CGLE3
has been proved in~\cite{Hone05} and~\cite{VernovCGLE} respectively. In
Section 4 we seek the elliptic solutions of the quintic complex
Ginzburg--Landau equation (CGLE5) using our modification of the Conte--Musette
method. Note that both the CGLE3 and the CGLE5 have only one-parameter
Laurent-series solutions in the wave form and there exist only a finite number
of such solutions.

Our approach can be effectively used in investigation of any system (\ref{f})
or a single differential equation, for which only a finite number of different
Laurent-series solutions exist. Note that for wide class of such differential
equations it has been proved that all their meromorphic solutions are elliptic
(maybe degenerated) functions~\cite{Eremenko}.

\section{Properties of the elliptic functions}

Let us recall some definitions and theorems.
 The function $\varrho(z)$ of the complex variable $z$ is
 a doubly-periodic function if
there exist two numbers $\omega_1$ and $\omega_2$ with $\omega_1/\omega_2
\not\in \mathbb{R}$, such that for all $z\in \mathbb{C} $
\begin{equation}
  \varrho(z)=\varrho(z+\omega_1)=\varrho(z+\omega_2).
\end{equation}

By definition a double-periodic meromorphic function is called an elliptic
function~\cite{BE,Hurwitz}. These periods define the period parallelograms
with vertices $z_0$, $z_0+N_1\omega_1$, $z_0+N_2\omega_2$ and
$z_0+N_1\omega_1+N_2\omega_2$, where $N_1$ and $N_2$ are arbitrary natural
numbers and $z_0$ is an arbitrary complex number. The fundamental
parallelogram of periods is called a parallelogram of period, which does not
include other parallelogram of periods, that corresponds to $N_1=N_2=1$.

The classical theorems for elliptic functions~\cite{BE,Hurwitz} prove that

\begin{itemize}

\item If an elliptic function has no poles then it is a constant.

\item The number of elliptic function poles within any finite period
parallelogram  is finite.

\item The sum of residues within any finite period parallelogram is equal to
zero (\textbf{the residue theorem}).

\item If $\varrho(z)$ is an elliptic function then any rational function of
$\varrho(z)$ and its derivatives is an elliptic function as well.

\item For each elliptic function $\varrho(z)$ there exist such $m$
($m\geqslant 2$) and such coefficients $h_{i,j}$ that $\varrho(z)$ is a
solution of Eq.~(\ref{subequ}).
\end{itemize}

\newtheorem{l1}{Lemma}

\begin{l1}

An elliptic function can not have two poles with the same Laurent series
expansions in its fundamental parallelogram of periods.

\textbf{Proof.}

{\rm Let some elliptic function $\varrho(\xi)$ has two poles in points $\xi_0$
and $\xi_1$, which belong to the fundamental parallelogram of periods. The
corresponding Laurent series are the same and have the convergence radius $R$.
Then the function $\upsilon(\xi)=\varrho(\xi-\xi_0)- \varrho(\xi-\xi_1)$ is an
elliptic function as a difference between two elliptic functions with the same
periods. At the same time for all $\xi$ such that $|\xi|<R$ \
$\upsilon(\xi)=0$, therefore, $\upsilon(\xi)\equiv 0$ and
$\varrho(\xi-\xi_0)\equiv\varrho(\xi-\xi_1)$ and $\xi_1-\xi_0$ is a period of
$\varrho(\xi)$. It contradicts to our assumption that both points $\xi_0$ and
$\xi_1$ belong to the fundamental parallelogram of periods.}
\end{l1}

\section{Construction of elliptic solutions}
\subsection{The quintic complex Ginzburg--Landau equation}

The one-dimensional quintic complex Ginzburg--Landau equation (CGLE5) is a
generalization of the one-dimensional cubic complex Ginzburg--Landau
equation~\cite{GL} (CGLE3), which is one of the most-studied nonlinear
equations (see~\cite{review} and references therein). Moreover, the CGLE5 is a
generic equation which describes many physical phenomena, for example, the
behaviour of travelling patterns in binary fluid convection~\cite{vSH} and the
large-scale behavior of many nonequilibrium pattern-forming systems~\cite{ch}.

The CGLE5 is as follows
\begin{equation}
\mathrm{i} \mathcal{A}_t + p \mathcal{A}_{xx} + q |\mathcal{A}|^2 \mathcal{A}
+r|\mathcal{A}|^4 \mathcal{A} - \mathrm{i} \gamma \mathcal{A} =0,
\label{eqCGL5}
\end{equation}
 where subscribes denote partial derivatives: $\mathcal{A}_t\equiv\frac{\partial \mathcal{A}}{\partial
t}$,  $\mathcal{A}_{xx}\equiv\frac{\partial^2 \mathcal{A}}{\partial x^2}$,
$p,q,r\in \mathbb{C}$ and $\gamma \in \mathbb{R}$.

One of the most important directions in the study of the CGLE5 is the
consideration of its travelling wave
reduction~\cite{Akhmediev,CoMu04,Doelman,Conte93,Saarloos,vSH}:
\begin{equation}
\mathcal{A}(x,t)=\sqrt{M(\xi)} \e^{\displaystyle
\mathrm{i}(\varphi(\xi)-\omega t)},\quad \xi=x-ct,\qquad c\in \mathbb{R},
\quad\omega \in \mathbb{R}.\label{travwave}
\end{equation}

Substituting (\ref{travwave}) in (\ref{eqCGL5}) and multiplying both sides of
this equation on $4M^2/A$ we obtain
\begin{equation}
\begin{array}{l}
\displaystyle 2pM''M-p{M'}^2+4\mathrm{i}p\psi MM'+2\Bigl(2\omega
-\mathrm{i}c-2\mathrm{i}\gamma+{}\\[2.7mm]
{}+2c\psi-2p\psi^2+2\mathrm{i}p\psi'
 \Bigr)M^2+4qM^3+4rM^4=0,
\end{array}
\label{equ1}
\end{equation}
where $\psi\equiv\varphi'\equiv\frac{\d\varphi}{\d\xi}$, $M'\equiv\frac{\d
M}{\d\xi}$. Equation (\ref{equ1}) is a system of two equations: both real and
imaginary parts of its left-hand side have to be equal to zero. Dividing
(\ref{equ1}) by $p$ and separating real terms from imaginary ones, we obtain
the following system
\begin{equation}
\!\left\{
\begin{array}{l}
\displaystyle 2 MM''-{M'}^2 - 4M^2\tilde{\psi}^2-
2\tilde{c}MM'+4g_iM^2 + 4d_r M^3+4u_rM^4=0,\\[2.7mm]
\displaystyle M\tilde{\psi}' + \tilde{\psi}\left(M' - \tilde{c}M\right)-g_rM +
d_i M^2 + u_iM^3=0,
\end{array}
\right.
\label{SYSTEM}
\end{equation}
where new real variables are as follows
\begin{equation}
u_r+\mathrm{i}u_i=\frac{r}{p},\qquad d_r + \mathrm{i} d_i = \frac{q}{p},\qquad
s_r - \mathrm{i} s_i = \frac{1}{p},
\end{equation}
\begin{equation}
g_r + \mathrm{i} g_i = (\gamma + \mathrm{i} \omega)(s_r - \mathrm{i} s_i) +
\frac{1}{2} c^2 s_is_r+\frac{\mathrm{i}}{4}c^2s_r^2,
\end{equation}
and
\begin{equation}
\tilde\psi\equiv\psi-\frac{cs_r}{2}, \qquad \tilde{c}\equiv cs_i.
\end{equation}

System (\ref{SYSTEM}) includes seven numerical parameters: $g_r$, $g_i$,
$d_r$, $d_i$, $u_r$, $u_i$ and $\tilde{c}$. Note that to obtain (\ref{SYSTEM})
from (\ref{equ1}) we assume that the functions $M(\xi)$ and $\psi(\xi)$ are
real.

The standard way to construct  exact solutions for system (\ref{SYSTEM}) is to
transform it into the equivalent third order differential equation for $M$. We
rewrite the first equation of system (\ref{SYSTEM}) as
\begin{equation}
\label{phi2G}
    \tilde{\psi}^2=\frac{G}{M^2},
\end{equation}
where
\begin{equation}
\label{G} G\equiv \frac{1}{2} M M'' - \frac{1}{4} M'^2
  -\frac{\tilde{c}}{2} M M' + g_i M^2 + d_r M^3+u_r M^4.
\end{equation}
From (\ref{phi2G}) it follows that
\begin{equation}
\label{dphi2G}
   \tilde{\psi}'\tilde{\psi}=\frac{G'M-2GM'}{2M^3},
\end{equation}

 Multiplying the second equation of (\ref{SYSTEM}) on $\tilde{\psi}$ and substituting
(\ref{phi2G}) and (\ref{dphi2G}) in it, we express $\tilde{\psi}$ in terms of
$M$ and its derivatives:
\begin{equation}
 \tilde{\psi} = \frac{G'-2 \tilde{c}  G}{2M^2\left(g_r - d_i M-u_i M^2\right)}, \label{psi}
\end{equation}
and obtain the third order equation for $M$:
\begin{equation}
(G'-2\tilde{c}G)^2 + 4GM^2(g_r-d_i M - u_i M^2)^2=0. \label{Gequ}
\end{equation}

\subsection{The Laurent series solutions}

Below we consider the case
\begin{equation}
\frac{p}{r}\not\in \mathbb{R}, \label{pqnotR}
\end{equation}
which corresponds to the condition $u_i\neq 0$. In this case Eq.~(\ref{Gequ})
is not integrable~\cite{Conte93} and its general solution (which should depend
on three arbitrary integration constants) is not known. Using the Painlev\'e
analysis~\cite{Conte93} it has been shown that single-valued solutions of
(\ref{SYSTEM}) can depend on only one arbitrary parameter.
System~(\ref{SYSTEM}) is autonomous, so this parameter is $\xi_0$: if
$M=f(\xi)$ is a solution, then $M=f(\xi-\xi_0)$, where $\xi_0\in \mathbb{C}$
has to be a solution. Special solutions in terms of elementary functions have
been found in~\cite{Akhmediev,CoMu04,Conte93,Saarloos}. All known exact
solutions of (\ref{SYSTEM}) are elementary (rational or hyperbolic) functions.
The full list of these solutions is presented in~\cite{CoMu04}. The purpose of
this section is to find at least one elliptic solution of~(\ref{SYSTEM}).

System~(\ref{SYSTEM}) is invariant under the transformation:
\begin{equation}
    \tilde{\psi} \rightarrow {} - \tilde{\psi}, \quad  g_r \rightarrow {} -
    g_r,
    \quad  d_i \rightarrow {} - d_i, \quad  u_i \rightarrow {} - u_i,
\end{equation}
therefore we can assume that $u_i>0$ without loss of generality. Moreover,
using scale transformations:
\begin{equation}
    M \rightarrow \lambda M,  \quad  d_r \rightarrow \frac{d_r}{\lambda},
    \quad d_i \rightarrow {} \frac{d_i}{\lambda}, \quad  u_r\rightarrow
    \frac{u_r}{\lambda^2},  \quad  u_i\rightarrow
    \frac{u_i}{\lambda^2},
\end{equation}
we can always put $u_i=1$.

Let us construct the Laurent series solutions to system~(\ref{SYSTEM}). We
assume that in a sufficiently small neighborhood of the  singularity point
$\xi_0$ the functions $\tilde{\psi}$ and $M$ tend to infinity as some powers
of $\xi-\xi_0$:
\begin{equation}
\label{Leadterms} \tilde{\psi}=A(\xi-\xi_0)^\alpha\qquad \mbox{and}\qquad
M=B(\xi-\xi_0)^\beta,
\end{equation}
where $\alpha$ and $\beta$ are negative integer numbers and, of course, $A\neq
0$ and $B\neq 0$. Substituting~(\ref{Leadterms}) into~(\ref{SYSTEM}) we obtain
that two or more terms in the equations of system~(\ref{SYSTEM}) balance if
and only if $\alpha=-1$ and $\beta=-1$. In other words in this case these
terms have equal powers and the other terms can be ignored as
$t\longrightarrow t_0$. We obtain values of $A$ and $B$  from the following
algebraic system:
\begin{equation}
\label{ABsystem} \left\{
\begin{array}{l}
\displaystyle B^2\left(3-4A^2+4u_rB^2\right)=0,\\[2.7mm]
\displaystyle 2A-B^2=0.
\end{array}
\right.
\end{equation}

System (\ref{ABsystem}) has four nonzero solutions:
\begin{equation}
    A_1=u_r+\frac{1}{2}\sqrt{4u_r^2+3},\qquad
    B_1=\sqrt{2u_r+\sqrt{4u_r^2+3}},
\label{AB1}
\end{equation}
\begin{equation}
\label{AB2}
    A_2=u_r+\frac{1}{2}\sqrt{4u_r^2+3},\qquad
    B_2={}-\sqrt{2u_r+\sqrt{4u_r^2+3}},
\end{equation}
\begin{equation}
\label{AB3}
   A_3=u_r-\frac{1}{2}\sqrt{4u_r^2+3},\qquad
    B_3= \sqrt{2u_r-\sqrt{4u_r^2+3}}
\end{equation}
and
\begin{equation}
\label{AB4}
    A_4=u_r-\frac{1}{2}\sqrt{4u_r^2+3},\qquad
    B_4={}- \sqrt{2u_r-\sqrt{4u_r^2+3}}.
\end{equation}

Therefore, system (\ref{SYSTEM}) has four types of the Laurent series
solutions. Denote them as follows:
\begin{equation}
    \tilde{\psi}_k=\frac{A_k}{\xi}+a_{k,0}^{\vphantom{27}}+a_{k,1}^{\vphantom{27}}\xi+\dots, \qquad
    M_k=\frac{B_k}{\xi}+b_{k,0}^{\vphantom{27}}+b_{k,1}^{\vphantom{27}}\xi+\dots,
\label{Laura}
\end{equation}
where $k=1..4$.

Let $M(\xi)$ is a nontrivial elliptic function. Note that if $\tilde{\psi}$ is
a constant, then from the second equation of system~(\ref{SYSTEM}) it follows
that $M$ can not be a nontrivial elliptic function. Therefore, using
(\ref{psi}), we conclude that $\tilde{\psi}(\xi)$ has to be a nontrivial
elliptic function as well.

Let us consider the fundamental parallelogram of periods for the function
$M(\xi)$ and define a number of its poles in this domain. Let $M$ has a pole
of type $M_1$, hence, according to the residue theorem, it should has a pole
of type $M_2$ (it can not have a pole of type $M_4$ because $u_r$ is a real
parameter). So $\tilde{\psi}$ has poles with the Laurent series
$\tilde{\psi}_1$ and $\tilde{\psi}_2$. As an elliptic function it should have
a pole of type $\tilde{\psi}_3$ or $\tilde{\psi}_4$ as well. It means that the
function $M(\xi)$ should have a pole of type $M_3$ and, hence, a pole of type
$M_4$. So $M(\xi)$ should have at least four different poles in its the
fundamental parallelogram of periods. Using \textbf{Lemma 1}, we obtain that
the function $M(\xi)$ can not have the same poles in the fundamental
parallelogram of periods. Therefore, $M(\xi)$ has exactly four poles in its
fundamental parallelogram of periods. In this case by means of the residue
theorem for $\tilde{\psi}$ we obtain
\begin{equation}
\label{ur0}
    u_r=0.
\end{equation}

We obtain that the CGLE5 with $u_r\neq 0$ has no elliptic solution in the wave
form. In the case $u_r=0$ possible elliptic solutions should have four simple
poles in the fundamental parallelogram of periods, and, therefore, has the
following form~\cite{Hurwitz}:
\begin{equation}
\label{MMM}
    M(\xi-\xi_0)=C+\sum_{k=1}^4B_k\zeta(\xi-\xi_k),
\end{equation}
where the function $\zeta(\xi)$ is an integral of the Weierstrass elliptic
function multiplied by $-1$:
\begin{equation}
\zeta'(\xi)=-\wp(\xi),
\end{equation}
$C$ and $\xi_k$ are constants to be defined. We also should define periods of
the Weierstrass elliptic function.

To obtain restrictions on other parameters, we use the Hone's
method~\cite{Hone05} and apply the residue theorem to the functions
$\tilde{\psi}^2$, $\tilde{\psi}^3$, and so on. The residue theorem for the
function $\tilde{\psi}^2$ gives the equation:
\begin{equation}
\label{psi2equ}
    \sum\limits_{k=1}^4 A_ka_{k,0}^{\vphantom{27}}=0.
\end{equation}

The values of $a_{k,0}^{\vphantom{27}}$ are as follows ($u_r=0$):
\begin{equation}
   a_{1,0}^{\vphantom{27}}={\frac{\sqrt {3}}{48}}\left( 6\tilde{c}
   -\sqrt [4]{27}d_i-15\sqrt[4]{3}d_r\right),
\end{equation}
\begin{equation}
a_{2,0}^{\vphantom{27}}={\frac{\sqrt {3}}{48}}\left( 6\tilde{c}
   +\sqrt [4]{27}d_i+15\sqrt[4]{3}d_r\right),
\end{equation}
\begin{equation}
   a_{3,0}^{\vphantom{27}}={}-{\frac {\sqrt {3}}{48}}\left(6\tilde{c}
   +\mathrm{i}\left(\sqrt [4]{27}d_i- 15\sqrt
[4]{3}d_r\right)\right),
\end{equation}
\begin{equation}
   a_{4,0}^{\vphantom{27}}={}-{\frac{\sqrt {3}}{48}}\left(6\tilde{c}
   -\mathrm{i}\left(\sqrt [4]{27}d_i- 15\sqrt
[4]{3}d_r\right)\right).
\end{equation}

Substituting $A_k$ and $a_{k,0}$ in (\ref{psi2equ}), we obtain
\begin{equation}
    \sum\limits_{k=1}^4 A_ka_{k,0}^{\vphantom{27}}=\frac{3}{4}\tilde{c}=0,
\end{equation}
therefore $\tilde{c}=0$.

For the function $\tilde{\psi}^3$ the residue theorem gives:
\begin{equation}
\label{psi3equ}
    \sum\limits_{k=1}^4 A_k\left(A_ka_{k,1}^{\vphantom{27}}+a_{k,0}^2\right)=0.
\end{equation}
The values $a_{k,1}^{\vphantom{27}}$ are as follows (we put $\tilde{c}=0$):
\begin{equation}
 a_{1,1}^{\vphantom{27}}=\frac{1}{384}\left(3{{d_i}}^{2}-63{d_r}^2-66\sqrt{3}d_id_r+128\sqrt
{3}g_i\right),\qquad a_{2,1}^{\vphantom{27}}=a_{1,1}^{\vphantom{27}},
\end{equation}
\begin{equation}
 a_{3,1}^{\vphantom{27}}=\frac{1}{384}\left(3{d_i}^{2}-63{d_r}^2+66\sqrt{3}d_id_r-128\sqrt
{3}g_i\right),\qquad a_{4,1}^{\vphantom{27}}=a_{3,1}^{\vphantom{27}}.
\end{equation}

Equation (\ref{psi3equ}) is equivalent to
\begin{equation}
{d_i}^2+27{d_r}^2=0 \quad \rightarrow \quad d_i=\pm \mathrm{i}\sqrt{27}d_r.
\end{equation}

The parameters $d_r$ and $d_i$ should be real, therefore, $d_r=0$ and $d_i=0$.
So, consideration of $\tilde{\psi}^2$ and $\tilde{\psi}^3$ gives three
restrictions:
\begin{equation}
\label{ristrict2} \tilde{c}=0, \qquad d_r=0 \qquad \mbox{and} \qquad d_i=0.
\end{equation}

The residue theorem for $\tilde{\psi}^4$ gives the restriction
\begin{equation}
\label{gigr0}
g_ig_r=0.
\end{equation}

Considering $\tilde{\psi}^5$ and $\tilde{\psi}^6$, we do not obtain new
restrictions on coefficients. Taking into account (\ref{ur0}) and
(\ref{ristrict2}) we obtain system (\ref{SYSTEM}) in the following form:
\begin{equation}
\left\{
\begin{array}{l}
\displaystyle 2MM''-{M'}^2 - 4M^2\tilde{\psi}^2 +4g_iM^2=0,\\[2.7mm]
\displaystyle \tilde{\psi}'M + \tilde{\psi} M'-g_rM+M^3=0.
\end{array}
\right. \label{SYSTEM27}
\end{equation}

To find elliptic solutions to system (\ref{SYSTEM27}) we use the
Conte--Musette method. The function $\tilde{\psi}(\xi)$ can have not four but
two different Laurent series expansions, whereas the functions $M(\xi)$ should
have four different Laurent series expansions, so it is easier to find
$\tilde{\psi}(\xi)$ than $M(\xi)$. Equation~(\ref{subequ}) with $m=1$ has no
elliptic solution.  Let $\tilde{\psi}(\xi)$ satisfies Eq.~(\ref{subequ}) with
$m=2$:
\begin{equation}
\label{equpsi}
\begin{array}{l}
{\tilde{\psi}'{}}^2+\left(\tilde{h}_{2,1}\tilde{\psi}^2+\tilde{h}_{1,1}\tilde{\psi}
+\tilde{h}_{0,1}\right)\tilde{\psi}'
+{}\\{}+\tilde{h}_{4,0}\tilde{\psi}^4+\tilde{h}_{3,0}\tilde{\psi}^3
+\tilde{h}_{2,0}\tilde{\psi}^2+\tilde{h}_{1,0}\tilde{\psi}+\tilde{h}_{0,0}=0.
\end{array}
\end{equation}

Substituting in (\ref{equpsi}) the Laurent series of $\tilde{\psi}$, which
begins from $A_1$ (more exactly we use the first ten coefficients), we obtain
the following solution $\tilde{h}_{k,j}$ for an arbitrary value of the
parameter $g_r\neq 0$ and $g_i=0$:
\begin{equation}
\label{solC1}
 \tilde{h}_{4,0}={}-\frac{4}{3}, \  \tilde{h}_{0,0} ={}-\frac{g_r^2}{9},
\  \tilde{h}_{3,0}=\tilde{h}_{2,0}=\tilde{h}_{1,0}=\tilde{h}_{0,1}
=\tilde{h}_{1,1}=\tilde{h}_{2,1}=0,
\end{equation}
a few solutions with $g_i=0$ and $g_r=0$ and no solution for $g_i\neq 0$.

The straightforward substitution of the functions
\begin{equation}
    \breve{\psi}=\frac{\sqrt{3}}{2t},\qquad
    \breve{M}={}\pm\frac{\sqrt[4]{3}}{t},
\end{equation}
or
\begin{equation}
   \hat{\psi}={}-\frac{\sqrt{3}}{2t},\qquad
    \hat{M}={}\pm \mathrm{i}\frac{\sqrt[4]{3}}{t}
\end{equation}
in (\ref{SYSTEM27}) with $g_r=0$ and $g_i=0$ proves that they are exact
solutions. The coefficients of the Laurent-series solutions do not include
arbitrary parameters, so the obtained solutions are unique single-valued
solutions and system~(\ref{SYSTEM27}) has no elliptic solution for these
values of parameters.

In the case of solutions (\ref{solC1}) the function $\tilde{\psi}(\xi)$
satisfies the equation
\begin{equation}
{\tilde{\psi}'{}}^2=\frac{4}{3}\tilde{\psi}^4+\frac{g_r^2}{9}.
 \label{equel}
\end{equation}
The polynomial in the right hand side of (\ref{equel}) has four different
roots, therefore $\tilde{\psi}$ is a non-degenerate elliptic
function~\cite{Hurwitz}.

Surely we do not rigorously prove the existence of elliptic solutions to the
CGLE5. More precisely, we calculate fifty coefficients of the Laurent series
of the function $\tilde{\psi}$, which satisfies (\ref{SYSTEM27}) with $g_i=0$
and find that they coincide with corresponding coefficients of the Laurent
series of the exact solution to Eq.~(\ref{equel}).

For rigorous proof we should find the function $M(\xi)$ and check that this
function is a solution of $(\ref{Gequ})$. The function $M(\xi)$ in a
parallelogram of periods has four different Laurent series expansions, so we
should choose the parameter $m$ such that solutions of Eq.~(\ref{subequ}) have
four poles in its fundamental parallelogram of periods. Minimal possible value
of $m$ is equal to 4. The general form of (\ref{subequ}) for $m=4$ and $p=1$
is the following:
\begin{equation}
\begin{array}{@{}l@{}}
M'^4+\left(h_{2,3}M^2+h_{1,3}M+h_{0,3}\right)M'^3+{}\\{}+\left(h_{4,2}M^4+h_{3,2}M^3+h_{2,
2}M^2+h_{1, 2}M+h_{0,2}\right)M'^2+{}\\
{}+\left(h_{6,1}M^6+h_{5,1}M^5+h_{4, 1}M^4+h_{3, 1}M^3+h_{2,1}M^2+h_{1, 1}M+{}
\right.\\ \left. {}+h_{0,
1}\right)M'+h_{8,0}M^8+h_{7,0}M^7+h_{6,0}M^6+h_{5,0}M^5+{}\\{}+h_{4,0}M^4+h_{3,0}M^3+h_{2,
0}M^2+h_{1,0}M+h_{0,0}=0.
\end{array}
\label{equM}
\end{equation}

Substituting the Laurent series $M_k$ from (\ref{Laura}), we transform the
left hand side of (\ref{equM}) into the Laurent series, which has to be equal
to zero. Therefore, we obtain the algebraic system in $h_{i,j}$ and $g_r$. The
first algebraic equation, which corresponds to $1/\xi^8$ is
\begin{equation}
\label{equM1}
 B_k^4\left(h_{8,0}B_k^4-h_{6,1}B_k^3+h_{4,2}B_k^2-h_{2,3}B_k+1\right)=0,
\end{equation}
where $B_k$ is defined by (\ref{AB1})--(\ref{AB4}). If we would attempt to
find all elliptic and degenerate elliptic solutions then we should use only
one of $B_k$ and can express, for example, $h_{8,0}$ via $h_{6,1}$, $h_{4,2}$
and $h_{2,3}$. We seek only elliptic solutions, so we know that all $B_k$ have
to satisfy (\ref{equM1}) and can consider Eq. (\ref{equM1}) as the following
system:
\begin{equation}
\label{equM1s} \left\{
\begin{array}{l}
h_{8,0}B_1^4-h_{6,1}B_1^3+h_{4,2}B_1^2-h_{2,3}B_1+1=0,\\
h_{8,0}B_2^4-h_{6,1}B_2^3+h_{4,2}B_2^2-h_{2,3}B_2+1=0,\\
h_{8,0}B_3^4-h_{6,1}B_3^3+h_{4,2}B_3^2-h_{2,3}B_3+1=0,\\
h_{8,0}B_4^4-h_{6,1}B_4^3+h_{4,2}B_4^2-h_{2,3}B_4+1=0.\\
\end{array}
\right.
\end{equation}
Using the explicit values of $B_k$ from (\ref{AB1})--(\ref{AB4}), we obtain
that
\begin{equation}
h_{8,0}={}-\frac{1}{3},\quad h_{4,2}=0, \quad h_{6,1}=0, \quad h_{2,3}=0.
\label{h}
\end{equation}
Taking into account (\ref{h}), from other equations of the algebraic system we
obtain
\begin{equation}
     h_{6,0}= \frac{4}{3}g_r, \quad h_{4,0} ={}-\frac{16}{9}g_r^2, \quad
     h_{2,0} = \frac{64}{81}g_r^3,
\end{equation}
all other $h_{i,j}$ are equal to zero. So, the equation for $M$ has the form
\begin{equation}
\label{equM4}
    {M'}^4=\frac{1}{81}M^2\left(3M^2-4g_r\right)^3.
\end{equation}

Equation (\ref{Gequ}) at $u_i=1$, $u_r=0$,  $\tilde{c}=0$, $d_r=0$, $d_i=0$
and $g_i=0$ has the form:
\begin{equation}
\label{Gequ0}
\frac{1}{4}\left({M'''}\right)^2-\left(2MM''-{M'}^2\right)\left(M^2-g_r\right)^2=0.
\end{equation}

We multiply Eq.~(\ref{Gequ0}) on ${M'}^2$ and use Eq.~(\ref{equM4}) to express
all derivatives of $M(\xi)$ in terms of the function $M(\xi)$. The
straightforward calculation shows that any solution of (\ref{equM4}) satisfies
(\ref{Gequ0}). So, we obtain elliptic wave solutions of the CGLE5. If $s_i\neq
0$ these solutions are the standing wave solutions, in the opposite case
($s_i=0$) the solutions can have an arbitrary speed~$c$.

Note that we obtain (\ref{Gequ}) from (\ref{equ1}) using the condition that
$M(\xi)$ is a real function. For $g_r<0$ and any initial value of $M$ we
obtain real solutions. In the case $g_r>0$ there exists the minimal possible
initial value of $M$ for which real solutions  exist and only particular
solutions of (\ref{equM4}) are suitable elliptic solutions to the CGLE5. The
function $M(\xi)$ has the form (\ref{MMM}), the values of constants can be
determined from (\ref{equM4}).

Summing up we can conclude that our modification of the Conte--Musette method
allows us to get two results: we obtain new elliptic wave solutions of the
CGLE5, and we prove that these solutions are unique elliptic solutions for the
CGLE5 with $g_r\neq 0$.

From (\ref{gigr0}) it follows that elliptic solutions can exist if $g_r$ or
$g_i$ is equal to zero. For all nonzero values of $g_r$ and zero $g_i$ we have
found elliptic solutions. In the case $g_r=g_i=0$ there is no elliptic
solution. In the case of zero $g_r$ and nonzero $g_i$ we substitute the
obtained Laurent series solutions $M_k$ into Eq.~(\ref{subequ}) with
$m=1,\dots,4$ and do not obtain neither elliptic functions nor degenerate
elliptic solutions. We hope that the more detail analysis of this case allows
us to find all elliptic solutions for the CGLE5.

\section{Construct of elliptic solutions for nonintegrable systems}

The approach, which we have considered in the previous section, can be
applicable to many nonintegrable systems. The best applicability is
nonintegrable systems of autonomous nonlinear ordinary differential equations
with so-called finiteness property~\cite{Eremenko}: there is only a finite
number of formal Laurent series that satisfy the system. For such systems we
can propose the following way to the search for elliptic solutions:

\begin{enumerate}

\item  Calculate a few first terms of all solutions of system (\ref{f}) in the
form of the Laurent series.

\item Choose the function $y_k$, which should be elliptic. Check should other
functions be elliptic or not.

\item Using the residue theorem define values of numeric parameters at which
the solution $y_k$ can be an elliptic function.

\item Define a minimal number of poles for candidates to elliptic solutions.
Using this number, choose a positive integer $m$, define the form of
Eq.~(\ref{subequ}) and calculate the number of unknown coefficients $h_{j,k}$.

\item Calculate the sufficient number of coefficients for all Laurent series
of $y_k$ and substitute the obtained coefficients into Eq.~(\ref{subequ}).
This substitution transforms (\ref{subequ}) into a linear and over\-determined
system in $h_{j,k}^{\vphantom{27}}$ with coefficients depending on numerical
parameters.

\item  Eliminate coefficients $h_{j,k}^{\vphantom{27}}$ and get a nonlinear
system in parameters.

\item Solve the obtained nonlinear system.
\end{enumerate}

We restrict ourselves to the search of elliptic solutions only. This makes
possible to consider not one Laurent series, but as much as possible. This is
a key idea of our modification of the Conte--Musette method.

Using the Conte--Musette method in the original form, one in principal would
be able to find the elliptic solutions of the CGLE5, but calculations would be
cumbersome, because he should reobtain all known solutions in terms of
elementary functions to obtain new elliptic solutions. One has to construct
and to solve an algebraic system, which includes at least $32$ equations,
which are linear in $24$ unknowns $h_{i,j}$ and nonlinear in $7$ parameters.
In our approach, we seek only elliptic solutions and, using the Hone's method,
reduce the number of arbitrary numerical parameters to $1$. Consideration of a
few Laurent series, instead of one also simplifies the obtained system of
algebraic equations. We conclude that the use of the Conte--Musette method in
combination with the Hone's method is very effective.

In contrast to~\cite{Hone05,CoMu03} we consider a system of differential
equations instead of the equivalent single differential equation and
demonstrate that the system is more convenient for analysis than a single
equation for $M$. If we analyse only the Laurent series expansions of the
function $M$, then we should also consider the case when $M$ has not four, but
only two different Laurent series expansions, beginning from $B_1$ and $B_2$
or $B_3$ and $B_4$. The consideration of the Laurent series expansions of two
functions: $\tilde\psi$ and $M$, allows us to reject this possibility. Note
that in our opinion if one seek degenerate elliptic solutions using the
Laurent series then a system of equations is more convenient for analysis than
the equivalent single equation as well.

Moreover, in contrast to traditional methods when we use the Conte--Musette
method we can choose a function, which analytic form should be found. Instead
of the functions $\tilde\psi$ and $M$ we can consider some combination, for
example, a polynomial, of this functions and seek this combination in the
analytic form. Note that we have no need of any differential equation for this
combination. So, we can conclude that the use of the Laurent series solutions
gives new insight on problem of construction of exact solutions for
nonintegrable systems.

\section{Comparison with standard methods}

Let us compare our approach with other methods for construction of special
solutions of nonintegrable systems (see~\cite{CoMu04} as a review of these
methods).

Without the use of the Laurent series solutions it would be very difficult to
find the elliptic solutions of the CGLE5, because the form of Eq.~(\ref{equM})
is very complex. For example, the Fan's technology~\cite{Fan02,Fan03} allows
to find solutions of nonintegrable systems, which are polynomials of the
function $\varrho$, that satisfies the following equation:
\begin{equation}
    {\varrho'}^2=\sum_{j=0}^N c_j\varrho^j,
\end{equation}
where $N$ and $c_j$ are constants to be determined. The function $\tilde\psi$
is a solution of a similar equation with $N=4$, but, using the Fan's approach,
it is impossible to find $\tilde\psi(\xi)$ without knowledge of $M(\xi)$,
which satisfies the more complex equation~(\ref{equM}). The Kudryashov's
method~\cite{Kudryashov1} as well as the
methods~\cite{CoMu04,Fan02,Fan03,Pavlov,Kudryashov89,Kudryashov2,CoMu03,Nikitin,santos,Timosh99,VeTish04,VernovJNMP},
proposed to search both elliptic and elementary solutions, attempt to find
solutions for the initial differential system at all values of the numeric
parameters such that single-valued solutions can exist. The use of the residue
theorem and the Hone's method allows us to fix some of these parameters and to
simplify calculations without loss of elliptic solutions. Note that we not
only find elliptic solutions for the CGLE5, but also we prove that there are
no other elliptic solutions for $g_r \neq 0$. Using the standard methods one
can say nothing about uniqueness of the obtained special elliptic solutions.

To find solutions for the CGLE5 the authors of
papers~\cite{Akhmediev,Conte93,Saarloos} put some restrictions on dependence
between $\tilde\psi$ and $M$. The use of Laurent solutions allows us to search
$\tilde\psi$ without any restrictions and without eliminating $M$ from system
(\ref{SYSTEM}).

\section{Conclusions}
In this paper we propose a new approach for the search of elliptic solutions
to systems of differential equations. The proposed algorithm is a modification
of the Conte--Musette method~\cite{CoMu03}. We restrict ourselves to the
search of elliptic solutions only. A key idea of this restriction is to
simplify calculations by means of the use of a few Laurent series solutions
instead of one and the use of the residue theorem.

The application of our approach to the quintic complex one-dimensional
Ginz\-burg--Landau equation (CGLE5) allows to find elliptic solutions in the
wave form. Let us  point out that the obtained solutions are the first elliptic solutions
for the CGLE5. Using the investigation of the CGLE5 as an example, we
demonstrate that to find elliptic solutions the analysis of a system of
differential equations is more preferable than the analysis of the equivalent
single differential equation.

We also find restrictions on coefficients, which are necessary conditions for
the existence of elliptic solutions for the CGLE5. To do this we  develop the
Hone's method~\cite{Hone05}. We show that this method is useful not only to
prove the nonexistence of elliptic solutions, but also to find new elliptic
solutions. Note that the Hone's method and, therefore, our approach, are so
effective in the case of the CGLE5, because coefficients of the Laurent series
solutions depend only on parameters of equations, i.e. they do not include
additional arbitrary parameters (have no resonances). It is an important
problem to generalize the Hone's method on the Laurent series solutions with
resonances.

Another way for future investigations is the generalization of the
Conte--Musette method on the case of multivalued solutions. Some results in
this direction have been obtained in~\cite{VernovJNMP,Vernovprog}.

\section*{Acknowledgements}
 This work has been supported in part by Russian
Fede\-ration President's Grant NSh--1685.2003.2.


\end{document}